\documentclass[aps,prl,showpacs,twocolumn,amsmath,10pt,nofootinbib,superscriptaddress,floatfix]{revtex4}
\usepackage{graphicx,psfrag}
\usepackage{times}
\usepackage{color}
\usepackage{amsfonts,amssymb,stmaryrd,latexsym,amsmath}

\begin{document}
\title{Comment on ``Measurement of 2- and 3-nucleon short range correlation probabilities in nuclei''}
\author{Douglas W. Higinbotham}
\affiliation{Jefferson Lab, Newport News, VA 23606, USA}
\author{Or Hen}
\affiliation{Tel Aviv University, Tel Aviv, Israel}
\maketitle

Egiyan~{\it{et al.}}~\cite{Egiyan:2005hs} reported the first
observation of a 3-nucleon Short Range Correlation (SRC) plateau in
inclusive A/$^3$He $(e,e')$ ratios at $x_B = Q^2/2m\omega > 2$ at a
momentum transfer centered at 
$Q^2\approx1.6$ GeV$^2$; yet, a
subsequent measurement by Fomin~{\it{et al.}}~\cite{Fomin:2011ng} at
$Q^2=2.9$ GeV$^2$ did not reproduce the results.  While the
difference could be due to a $Q^2$ dependence, it would be
unexpected~\cite{Arrington:2011xs} especially since the two
measurements agreed in the $x_B < 2$ region.

The experiments used very different electron spectrometers.
Fomin~{\it{et al.}} used a small solid angle spectrometer with an
energy resolution, $\delta E/E \approx 10^{-3}$; while Egiyan~{\it{et
    al.}} used a large acceptance spectrometer with $\delta E/E
\approx 6\times10^{-3}$~\cite{Mecking:2003zu,Baghdasaryan}.  
While both experiments
presented their data as a function of $x_B$, they measured scattering electron
energies, E', to determine $\omega = E_{beam} - E'$ and $x_B$.  

\begin{figure}[ht!]
\begin{center}
\includegraphics[width=1.0\linewidth]{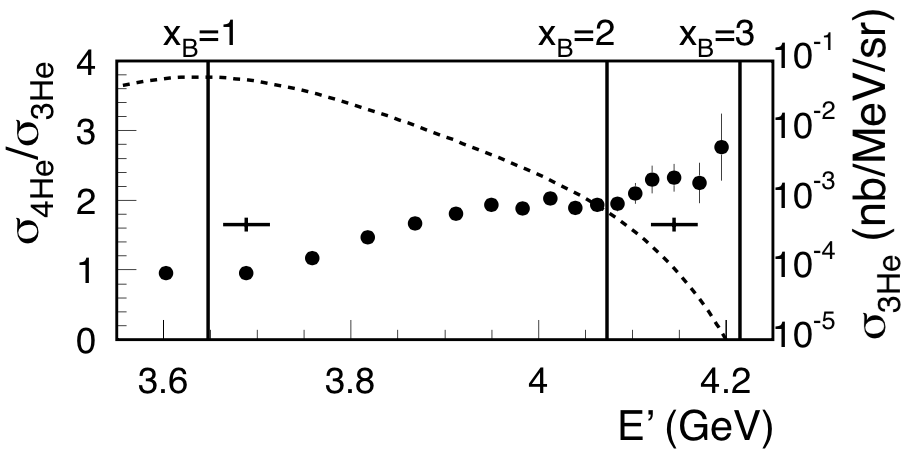}
\caption{The Egiyan cross section ratios plotted against the scattered electron energy, $E'$,
assuming an average $Q^2$ of 1.6~[GeV/c]$^2$.
The dashed-dotted curve shows the $^3$He(e,e') cross section~\cite{Zhihong}.   
The horizontal error bars at $E' = 3.68$ and 4.14 GeV show the $\pm0.6\%$ energy resolution.}
\label{fig:doug}
\end{center}
\end{figure}

Fig.~\ref{fig:doug}
shows the Egiyan~{\it{et al.}} $^4$He/$^3$He cross section ratios and
expected energy resolution $\sigma_{E'}=0.6\%$ as a function of $E'$ for a central $Q^2$ of
1.6~[GeV/c]$^2$.  
The data points were all placed at the center of each
bin.  The energy resolution is smaller than the bin spacing at small
$E'$ ($x_B\approx 1$)
but significantly larger than the bin spacing at large $E'$ ($x_B\approx 2$).

Fig.~\ref{fig:doug} shows that the $^3$He cross
section~\cite{Zhihong} at $x_b>2$ is also decreasing
very rapidly.  
Small bins where a cross section is decreasing rapidly, especially near a kinematic end point,
can be susceptible to a large fraction of events migrating from one bin to another~\cite{Lafferty:1994cj}.

\begin{figure}[ht!]
\begin{center}
\includegraphics[width=0.85\linewidth]{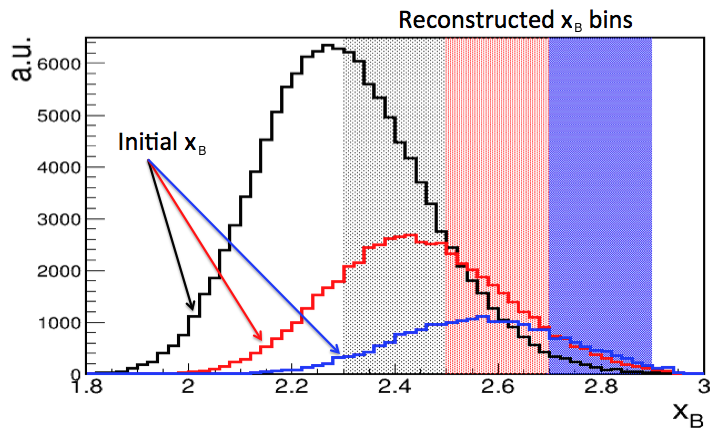}
\caption{(Color online) The vertical bands show bins that were used for
  three largest $x_B$ values in Egiyan {\it et al.} while the
  corresponding histograms
  show the original $x_B$ of the events populating those bins.
  This simulation show the $^3$He cross sections points at $x_B$ 2.40, 2.60 and 2.80 
  have only 37\%, 29\% and 20\% data that from within cut thus bin migration 
  would shift them to $x_B$ 2.30, 2.44 and 2.56 respectively.}
\label{fig:or}
\end{center}
\end{figure}

To test if this could be the source of the discrepancy,
we performed a Monte Carlo simulation
that generated electron scattering events at $Q^2=1.6$ GeV$^2$ based
on the $^3$He cross section shown in Fig.~\ref{fig:doug} and then smeared $E'$
for each event with an energy resolution of $\sigma=0.6\%$. 
The results, shown in Fig.~\ref{fig:or}, show how combining a decreasing cross-section with 
the moderate resolution can create large bin-migration effects
and how most of the events within the reconstructed $x_B$ bin were
likely from lower initial $x_B$'s.      
 
Thus, we find that the Egiyan {\it et al.} results at $x_B > 2$
are subject to large bin migration effects that, along with any
backgrounds, need to be taken into account before taking a ratio. 
We also note that by checking unphysical regions, such as $x_B > 2$
deuterium data, the magnitude of these undesired effects can be
experimentally determined.

We acknowledge many useful discussions with 
R.~Ent, N.~Fomin, S.~Stepanyan, L.~Weinstein, and Z.~Ye,
and the support of the Israel Science Foundation and 
Department of Energy Contract DE-AC05-060R23177.


\end{document}